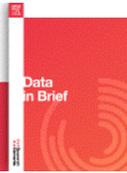

# ARTICLE INFORMATION

**Authors**

*Heitor Lifsitch[a], Gabriel Rocha[c], Hendrio Bragança[b], Cláudio Filho[c], Leandro Okimoto*[\*b], Allan Amorin[a], Fábio Cardoso[c]*

**Affiliations**

[a] *Department of R&D, TPV company, Amazonas, Brazil*

[b] *Institute of Computing, Federal University of Amazonas, Amazonas, Brazil*

[c] *Department of Electrical Engineering, State University of Amazonas, Amazonas, Brazil*

**Corresponding author's email address and Twitter handle**

*E-mail address: okimoto@icomp.ufam.edu.br*

**Abstract**

*To enhance the field of continuous motor health monitoring, we present FAN-COIL-I, an extensive vibration sensor dataset derived from a Fan Coil motor. This dataset is uniquely positioned to facilitate the detection and prediction of motor health issues, enabling a more efficient maintenance scheduling process that can potentially obviate the need for regular checks. Unlike existing datasets, often created under controlled conditions or through simulations, FAN-COIL-I is compiled from real-world operational data, providing an invaluable resource for authentic motor diagnosis and predictive maintenance research. Gathered using a high-resolution 32KHz sampling rate, the dataset encompasses comprehensive vibration readings from both the forward and rear sides of the Fan Coil motor over a continuous two-week period, offering a rare glimpse into the dynamic operational patterns of these systems in a corporate setting. FAN-COIL-I stands out not only for its real-world applicability but also for its potential to serve as a reliable benchmark for researchers and practitioners seeking to validate their models against genuine engine conditions.*

# SPECIFICATIONS TABLE

| | |
|---|---|
| **Subject** | *Computer Science.* |
| **Specific subject area** | *Data Mining and Machine Learning.* |
| **Type of data** | Processed, Analysed, Filtered. |
| **Data collection** | The dataset comprises two numpy arrays. The main array is shaped (5246, 320000, 2), indicating 5246 samples of vibration data at a 32KHz sampling rate from two sensor channels (forward and rear) of the Fan Coil motor. The second array, shaped (5246,), holds corresponding timestamps. The data was captured every 2 minutes using high-resolution vibration sensors on both the forward and rear sides of a Fan Coil motor over a period of 17 days. |
| **Data source location** | The dataset originated from a Fan Coil motor in a real-world corporate environment in the Manaus industrial district, localized in the state of Amazonas – Brazil. |
| **Data accessibility** | Data hosted in public repository [2].<br><br>Repository name:<br><br>**Vibration Sensor Dataset for Estimating Fan Coil Motor Health**<br><br>Data identification number: https://doi.org/10.6084/m9.figshare.25959403.v1<br><br>Direct URL to data:<br><br>https://figshare.com/articles/dataset/_i_Vibration_Sensor_Dataset_for_Estimating_Fan_Coil_Motor_Health_i_/25959403 |
| **Related research article** | None |

# VALUE OF THE DATA

- The FAN-COIL-I dataset is a data collection from a Fan Coil motor in a real industrial environment, providing information of motor's condition. This real-world applicability ensures that predictive maintenance models and health diagnostics developed using this dataset are practical, reliable, and ready for real-world implementation.
- The dataset provides a detailed perspective on motor operation through continuous data collecting for almost two weeks at a high sampling rate of 32KHz from both forward and rear sensors. The 32KHz sample rate enables the collection of even the smallest vibrations, facilitating an in-depth analysis of motor action. Having a high level of detail is essential for accurately identifying distinct fault patterns and enhancing the precision of maintenance models.
- The dataset has a temporal granularity that enables the examination of vibration patterns over time. Data was taken every 2 minutes for a period of 17 days. This provides for information about the temporal dynamics of motor health. The level of detail in this data is crucial for creating accurate models that can anticipate and prevent breakdowns, allowing for rapid maintenance.
- The collection captures data from both the forward and rear sides of the motor, providing a comprehensive perspective on the motor's operational condition. This dual view enables the ability to analyse in a more detailed and advanced manner, leading to improved accuracy in health diagnostics and predictive maintenance models.
- The dataset extended duration and comprehensive level of information facilitate the continuous evaluation of Fan Coil motors' condition, allowing for the timely identification of possible malfunctions. Adopting a long-term perspective is important for transitioning from a fixed maintenance schedule to a predictive maintenance approach, resulting in a substantial decrease in both downtime and maintenance expenses.
- A practical benchmark for research using the FAN-COIL-I offer an original standard for evaluating and verifying the accuracy of motor health monitoring and predictive maintenance models under real engine circumstances. This will significantly improve research in the field of continuous motor health monitoring, establishing an innovative benchmark for the quality and suitability of datasets.
- By providing a clear and structured format, FAN-COIL-I facilitates ease of access and analysis, empowering researchers and practitioners to explore innovative approaches to motor health monitoring.

# BACKGROUND

Motor anomalies are a significant concern in various industries due to their potential to cause unexpected failures and substantial downtime. Detecting these anomalies early is crucial for preventive maintenance, operational efficiency, safety, and cost savings. Vibration sensors play a vital role in this detection process, as motor vibration is influenced by numerous variables, making it one of the most useful data sources for identifying anomalies. Integrating vibration-based anomaly detection systems is essential for maintaining the reliability, efficiency, and safety of industrial operations, underscoring the critical need for advanced monitoring and diagnostic techniques [1].

# DATA DESCRIPTION

To accelerate the advances in the field of continuous motor health monitoring, we present the FAN-COIL-I dataset, a collection created to encourage advancements in predictive maintenance and health diagnostics of Fan Coil motors. This dataset is notable for its original source, detailed data, and wide range of operational patterns, making it an essential tool for study and development in the field of industrial machinery maintenance.

The FAN-COIL-I dataset was collected from a single Fan Coil motor, providing operating characteristics and health indications. The participants in this data collection consisted of the motor itself and the operational environment surrounding it. This allowed for a direct representation of real-world conditions, free from any human bias.

- A WEG[1] motor is mounted internally to a Fan Coil equipment, as shown in the following Figure 1a.
- The lifespan of this WEG motor ranges from 20 to 25 years, given proper and regular maintenance.
- The two-week period is sufficient to capture motor anomalies and normal behaviors.
- The fan coil is mounted on rubber pads immediately on the concrete floor.
- The fan coil is made of thin (deformable) metal sheets in a rectangular prism shape.
- The fan coil electric motor is mounted on the floor (metal sheet) of the equipment, as shown in the Figure 1b.
- Although the fan coil is mounted on a rigid surface (concrete base), the internal motor is mounted on a metal plate above the rubber cushion.
- The stiffness of the engine mounting surface is significantly lower than the stiffness of the concrete base below. Thus, we can consider that the engine is mounted on a flexible surface.
- The flexible surface promotes a greater amplitude vibration when compared to the rigid surface.
- The dataset includes vibration measurements from both the forward and rear sides of the motor, as shown in Figure 1c, collected continuously for a duration of 17 days. The prolonged duration of data collecting, along with a high sample rate of 32KHz for 10 seconds, offers a comprehensive and detailed understanding of the motor's performance and condition.
- The accelerometer sensor (786A) specification can be found in specification[2] document.

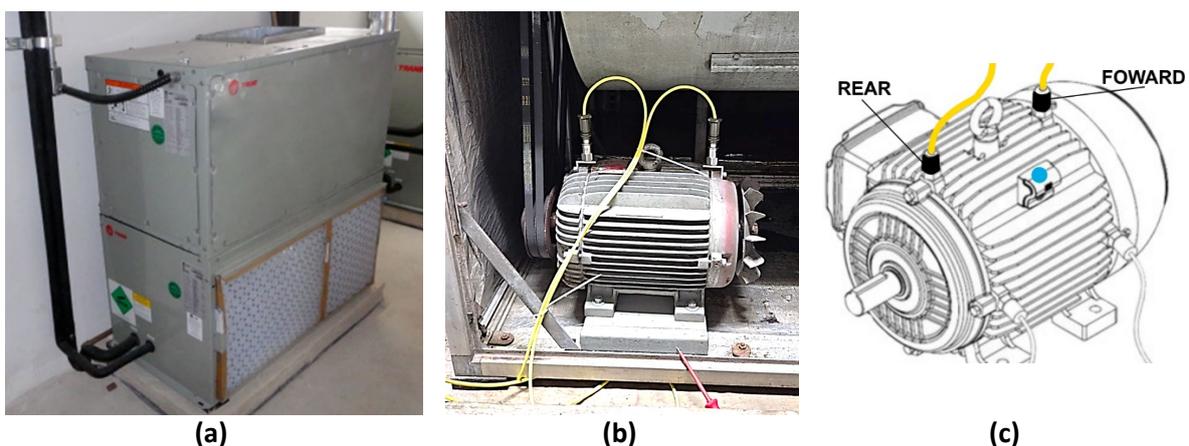

(a) (b) (c)

*Figure 1. Fan Coil motor setup.*

---

[1] https://static.weg.net/medias/downloadcenter/h50/h86/WEG-motor-scan-vibration-measurement-technical-note-10008451145-pt-en-es.pdf

[2] https://wilcoxon.com/wp-content/uploads/2022/11/786A_spec_98692E.2.pdf

The sensors collected data at a 2-minute interval for 10 seconds, with 5246 individual samples that offer a comprehensive overview of the motor's performance across varying operational conditions. The high-frequency data collection at 32KHz enables a detailed examination of vibration patterns, critical for identifying early signs of wear or failure. The dual-channel approach enhances the dataset's utility for developing predictive maintenance models that can accurately diagnose potential issues from multiple data perspectives.

In more detail, the dataset is composed of two numpy arrays. The primary array, shaped (5246, 320000, 2), captures vibration data across two channels, representing the motor's forward and rear, as shown in Figure 2. The second array contains timestamps for each of the 5246 samples, offering insights into the temporal aspects of the data. Table 1 presents the statistical analysis of the number of instances by sensors on FAN-COIL-I dataset. This structure allows for a granular analysis of vibration patterns, essential for diagnosing potential issues and predicting maintenance needs.

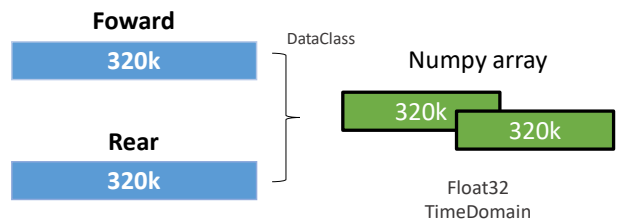

Figure 2. Processed array from Forward/Rear data class to numpy array.

Table 1: Statistical analysis of the number of instances by sensors on FAN-COIL-I dataset.

|           | Forward (Channel 0) | Rear (Channel 1) |
|-----------|---------------------|------------------|
| Mean      | -4.30271e-08        | 3.6558079e-06    |
| Std. Dev. | 0.178946            | 1.445250         |
| Min.      | -14.670534          | -14.72572        |
| 25%       | -0.088954           | -0.859762        |
| 50%       | 0.004544            | 0.000541         |
| 75%       | 0.092457            | 0.914099         |
| Max.      | 7.350223            | 10.305142        |

# EXPERIMENTAL DESIGN, MATERIALS AND METHODS

**DATA EXTRACTION**

To obtain the vibration readings presented in the FAN-COIL-I dataset, as shown in Figure 3, we follow a straightforward three-step process to transform basic sensor data into a pre-processed dataset:

1. Data Extraction: first, we use boards in the fan coil motor to collect real-time vibration data while it is operating. This data is transmitted in real-time to the server via the OPC-UA protocol and integrated into a basic RAMI-4.0 structure. This step ensures the creation of a digital twin for the fan coil motor's information on the server.

2. Labelling and standardize: the collected data are then labeled as either normal or abnormal, based on reports of the fan coil motor's normal functioning by an employee. We also standardize the data, as the sensor is sensitive to electric current. To ensure high precision, we normalize the data to G-force, maintaining a close relationship with the dataset.
3. Consolidation and cleaning: Finally, we process the RAMI-4.0 data class to convert it into a more commonly used data format for researchers. We use NumPy for this high-precision data, transforming the data class into NumPy format. This completes our data-cleaning process and removes the RAMI-4.0 structures.

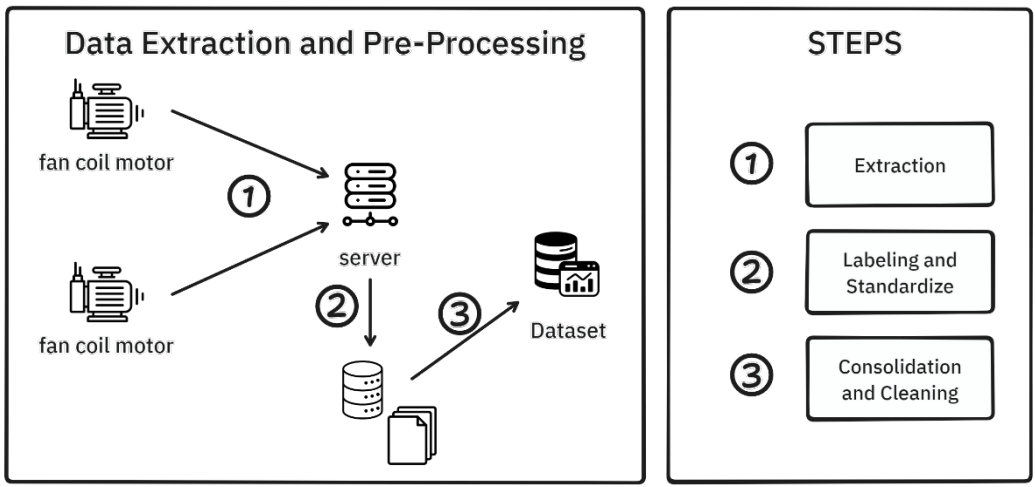

*Figure 3. Data extracting and pre-processing steps applied to create the FAN-COIL-I dataset.*

**ANALISYS PERFORMED WITH THE SENSOR**

We analyzed the dataset data using parameters found in its documentation. To do this, we need to convert the acceleration data to velocity. To obtain velocity from acceleration, we perform a numerical integration (cumulative sum) over time. Given an acceleration $a(t)$ sampled at a rate $f_s$:

$$v(t) = \int a(t).dt$$

*In discrete terms, the integration (cumulative sum) can be represented as:*

$$v[n] = \sum_{k=0}^{n} a[k].\Delta t$$

where, $v[n]$ is the velocity at the n-th sample, $a[k]$ is the acceleration at the k-th sample and $\Delta t = 1/f_s$ is the time interval between samples (sampling period). To convert from m/s to mm/s we multiply velocity by 1000.

The RMS value of a set of velocities $v_i$ over $N$ samples is given by:

$$RMS(v) = \sqrt{\frac{1}{N}\sum_{i=1}^{N} v_i^2}$$

We integrate acceleration to obtain velocity and converts it from meters per second (m/s) to millimeters per second (mm/s).

After obtaining the speed data, we compare the results obtained from the speed signals with the table provided in the engine documentation. We use power < 300kW with a flexible base (installation location). Figure 4 shows the differences between the forward and the rear sensors.

| RMS Vibration Speed [mm/s] | Power ≤ 300 kW Group 2 of ISO 10816-3 | | Power > 300 kW Group 1 of ISO 10816-3 | | | Labels | |
|---|---|---|---|---|---|---|---|
| | Rigid Base | Flexible Base | Rigid Base | Flexible Base | | | |
| V ≤ 2.8 | | | | | | | NORMAL |
| 2.8 < V ≤ 5.6 | | | | | | | ALERT |
| 5.6 < V ≤ 8.9 | | | | | | | CRITICAL |
| 8.9 < V ≤ 13.8 | | | | | | | |
| V > 13.8 | | | | | | | |

*Figure 4. Vibration alert and critical level limits – ISO 10816-3.*

Figure 5 shows that the RMS velocity data indicates fluctuating operating conditions for the system. There are periods of stability with lower RMS values, interrupted by intervals of increased stress where the values exceed the warning and occasionally the critical thresholds. The thresholds provide markers for evaluating the system's health, with frequent breaches into the warning range and occasional entries into the bad health range, indicating the need for closer monitoring or maintenance during those periods.

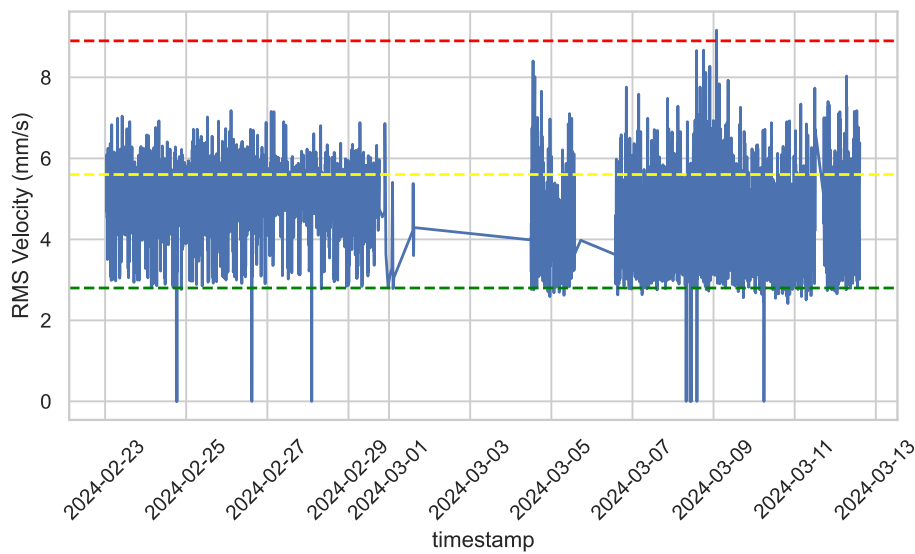

*Figure 5. Chart of the RMS values over time for forward sensor.*

The RMS velocity data for rear sensor, presented in Figure 6, now shows a clearer pattern of fluctuating operating conditions. Periods of stability with lower RMS values are interrupted by intervals of high stress, with RMS values often exceeding the warning and critical thresholds. This indicates the need for closer monitoring and possible intervention to prevent damage or failure during high-stress periods. The system seems to oscillate between states of high stress and lower stress, indicating potential issues with operational consistency or external factors influencing performance.

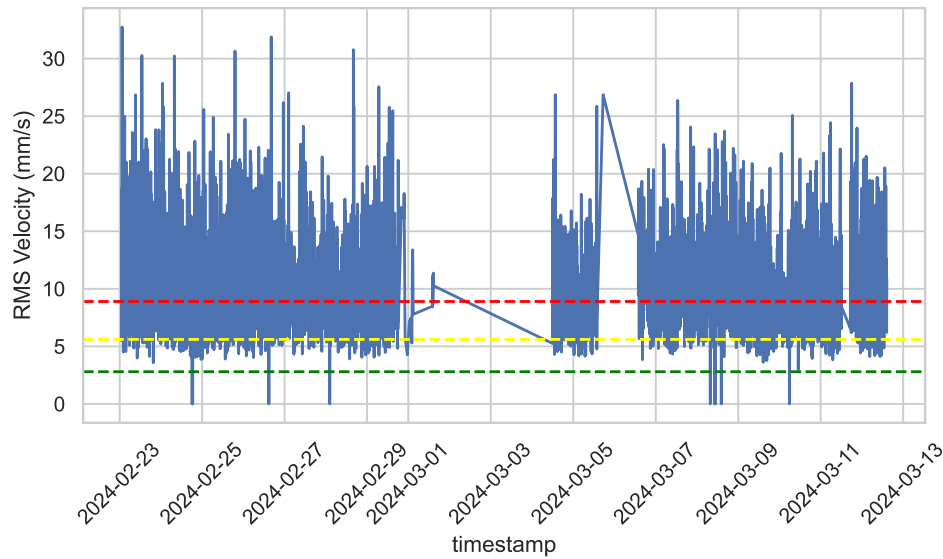

*Figure 6. Chart of the RMS values over time for rear sensor.*

To summarize, there is a notable disparity in the vibration levels recorded by the rear accelerometer (located near the pulley) and the forward accelerometer. The increased root mean square (RMS) velocity values might be linked to many underlying reasons. Although the vibration patterns and distances are similar, the vibration behavior differs greatly due to variations in geometry and the presence of different components in the system. The pulley on the back side may not be intrinsically misaligned, but its visibly skewed configuration results in elevated vibration levels on the rear side in comparison to the forward side. This discrepancy becomes more severe by the adaptable foundation of the system, which enables a more significant propagation of vibrations.

In addition, issues associated with the bearings, misalignment, and pulley looseness are contributing factors to the elevated RMS velocity values that have been found. Issues with the bearings, such as deterioration or damage, can have a substantial effect on the system's functioning, resulting in irregular patterns of vibration. System components that are not properly aligned, especially the pulley, also lead to higher levels of stress and vibration. Moreover, if there is slack in the pulley or other components, it allows for greater freedom of movement, leading to irregular and higher levels of vibration.

In order to address these problems, it is crucial to conduct a thorough examination and maintenance of the system's components. A careful inspection, with particular attention to the bearings and pulleys, should be performed to detect any indications of deterioration, misalignment, or slackness. Performing routine maintenance and realigning these components regularly will effectively decrease vibration levels and improve the overall operation of the system. It is important to consider making structural improvements, such as strengthening the system's base to decrease its flexibility, in order to lessen the transmission of vibrations.

Furthermore, it is essential to continuously monitor the root mean square (RMS) velocity and other pertinent data in order to promptly identify any issues. Utilize sophisticated diagnostic tools and procedures to evaluate vibration patterns and pinpoint the exact sources of problems. By identifying and resolving these fundamental causes and executing the suggested measures, it is feasible to attain a system functioning that is more consistent and dependable. Consistent monitoring and maintenance are crucial for guaranteeing optimal long-term performance and preventing any faults.

## LIMITATIONS

Some considerations regarding the conservation status of the Fan Coil electric motor that cannot be discarded are: i) there are no fan coil maintenance records; ii) there is no history of bearing changes, lubrication or alignment of pulleys, tensioning of pulleys; iii) there is dirt, particulates and moisture inside the equipment; iv) The condition of the centrifugal fan rotors was not checked.

## ETHICS STATEMENT

*The work did not involve any human subject or animal experiments.*

## CRediT AUTHOR STATEMENT

*Heitor Lifsitch: Methodology, Investigation, Data curation, Writing – review & editing. Gabriel Rocha: Conceptualization, Software, Methodology, Investigation, Data curation, Formal analysis. Hendrio Bragança: Conceptualization, Supervision, Writing – review & editing. Cláudio Filho: Conceptualization, Software, Supervision, Writing – review & editing. Leandro Okimoto: Conceptualization, Supervision, Formal analysis, Writing – review & editing. Allan Amorim Conceptualization, Supervision, Writing – review & editing. Fábio Cardoso: Conceptualization, Supervision, Writing – review & editing.*

## ACKNOWLEDGEMENTS

This research was financed in part by *Superintendência da Zona Franca de Manaus* (SUFRAMA) through the Informatics Law. The Informatics Law No. 8.387/91 (and amendments) in the Amazon region mandates that all companies producing IT goods and services annually invest at least 5% (five percent) of their gross domestic revenue.

## DECLARATION OF COMPETING INTERESTS

The authors declare that they have no known competing financial interests or personal relationships that could have appeared to influence the work reported in this paper.

## DECLARATION OF GENERATIVE AI AND AI-ASSISTED TECHNOLOGIES IN THE WRITING PROCESS

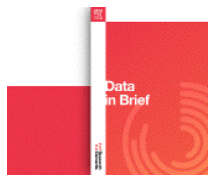

During the preparation of this work the author(s) used Grammarly, GoogleTranslator, CHATGPT to improve language and readability. After using this tool/service, the author(s) reviewed and edited the content as needed and take(s) full responsibility for the content of the publication.